\documentclass{article}

\usepackage{arxiv}

\usepackage[utf8]{inputenc} 
\usepackage[T1]{fontenc}    
\usepackage{hyperref}       
\usepackage{url}            
\usepackage{booktabs}       
\usepackage{amsfonts}       
\usepackage{nicefrac}       
\usepackage{microtype}      
\usepackage{lipsum}
\usepackage{graphicx}
\usepackage[utf8]{inputenc} 
\usepackage[T1]{fontenc}    
\usepackage{hyperref}       
\usepackage{url}            
\usepackage{booktabs}       
\usepackage{amsfonts}       
\usepackage{nicefrac}       
\usepackage{microtype}      
\usepackage{xcolor}         
\usepackage{graphicx}
\usepackage{subfigure}
\usepackage{amsmath}
\usepackage{amsthm}
\usepackage{amssymb}
\usepackage{enumitem}
\usepackage{adjustbox}
\usepackage{wrapfig}
\usepackage[normalem]{ulem}
\usepackage{mathrsfs}
\usepackage{multirow}
\usepackage{svg}

\usepackage{algorithm}
\usepackage{algorithmic}
\theoremstyle{definition}

\newcommand\algorithmicprocedure{\textbf{function}}
\newcommand{\algorithmicendprocedure}{\algorithmicend\ \algorithmicprocedure}
\makeatletter
\newcommand\FUNCTION[3][default]{%
  \ALC@it
  \algorithmicprocedure\ \textsc{#2}(#3)%
  \ALC@com{#1}%
  \begin{ALC@prc}%
}
\newcommand\ENDFUNCTION{%
  \end{ALC@prc}%
  \ifthenelse{\boolean{ALC@noend}}{}{%
    \ALC@it\algorithmicendprocedure
  }%
}
\newenvironment{ALC@prc}{\begin{ALC@g}}{\end{ALC@g}}

\graphicspath{ {./figures/} }

\newcommand{\NetBench}{\texttt{NetBench}}

\title{\NetBench: A Large-Scale and Comprehensive Network Traffic Benchmark Dataset for Foundation Models}

\begin{document}

\author{
  Chen Qian\footnotemark[1]\\
  Department of Computer Science \\
  William \& Mary \\
  \texttt{cqian03@wm.edu} \\
  \And
  Xiaochang Li\footnotemark[1] \\
  Department of Computer Science \\
  William \& Mary \\
  \texttt{xli59@wm.edu} \\
  \And
  Qineng Wang \\
  Independent Researcher \\
  \texttt{wangqineng73@gmail.com} \\
  \And
  Gang Zhou \\
  Department of Computer Science \\
  William \& Mary \\
  \texttt{gzhou@wm.edu} \\
  \And
  Huajie Shao \\
  Department of Computer Science \\
  William \& Mary \\
  \texttt{hshao@wm.edu} \\
}
\footnotetext[1]{These authors contributed equally to this work.}
\maketitle
\begin{abstract}

In computer networking, network traffic refers to the amount of data transmitted in the form of packets between internetworked computers or Cyber-Physical Systems. Monitoring and analyzing network traffic is crucial for ensuring the performance, security, and reliability of a network. However, a significant challenge in network traffic analysis is to process diverse data packets including both ciphertext and plaintext. While many methods have been adopted to analyze network traffic, they often rely on different datasets for performance evaluation. This inconsistency results in substantial manual data processing efforts and unfair comparisons. Moreover, some data processing methods may cause data leakage due to improper separation of training and testing data. To address these issues, we introduce the \NetBench, a large-scale and comprehensive benchmark dataset for assessing machine learning models, especially foundation models, in both network traffic classification and generation tasks. \NetBench~is built upon seven publicly available datasets and encompasses a broad spectrum of 20 tasks, including 15 classification tasks and 5 generation tasks. Furthermore, we evaluate eight State-Of-The-Art (SOTA) classification models (including two foundation models) and two generative models using our benchmark. The results show that foundation models significantly outperform the traditional deep learning methods in traffic classification. We believe \NetBench~will facilitate fair comparisons among various approaches and advance the development of foundation models for network traffic. Our benchmark is available at \url{https://github.com/WM-JayLab/NetBench}.





\end{abstract}


\section{Introduction}
\begin{figure}[!t]
    \centering
    \includegraphics[width=0.65\textwidth]{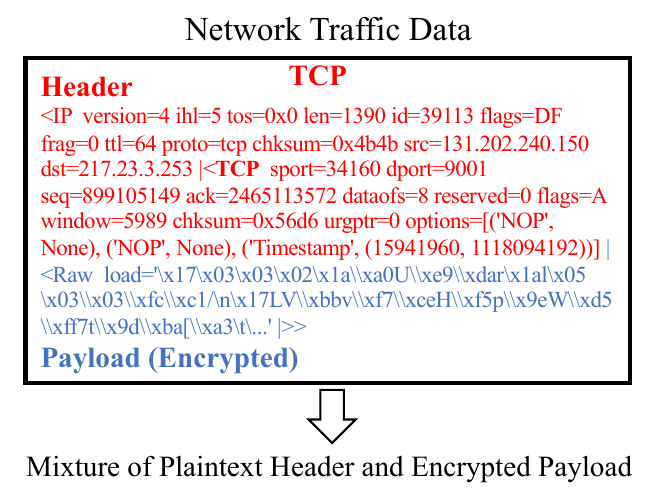}
    \caption{The format of network traffic data, which are the mixture of plaintext header and encrypted payload. This mixture makes it hard to directly process with text tokenizers for model training.}
    \label{fig:data_difference}
\end{figure}

In the domain of computer networking, network traffic~\cite{oliveira2016computer} is the amount of data transmitted in the form of packets between interconnected computers or systems. Generally, a network packet is composed of two parts: a header containing meta features and a commonly encrypted payload, as shown in Fig. \ref{fig:data_difference}. Analyzing network traffic is critical for enhancing network security and management. However, it is challenging to analyze network traffic due to the diversity of packet types, such as TCP and UDP, as well as the presence of both encrypted and unencrypted data. 




\begin{table*}[!t]
\centering
\caption{The statistical information of 7 different datasets, including 15 traffic classification tasks and 5 traffic generation tasks.}
\renewcommand\arraystretch{0.4}
\begin{tabular}{c | c | c | l | c}
\toprule
Dataset                                        & \#Flow                      & \#Packet                     & Task                                       & \#Label             \\
\midrule
\multirow{3}{*}{ISCXVPN 2016~\cite{vpn}}                  & \multirow{3}{*}{311,390}   & \multirow{3}{*}{1,040,354}  & 1 - VPN Detection                          & 2                  \\
                                               &                            &                             & 2 - VPN Service Detection                  & 6                  \\
                                               &                            &                             & 3 - VPN Application Classification         & 17                 \\
\midrule
\multirow{2}{*}{ISCXTor 2016~\cite{tor}}                  & \multirow{2}{*}{55,523}    & \multirow{2}{*}{1,163,495}  & 4 - Tor Detection                          & 2                  \\
                                               &                            &                             & 5 - Tor Service Detection                  & 7                  \\
\midrule
\multirow{2}{*}{USTC-TFC 2016~\cite{ustc}}                 & \multirow{2}{*}{489,139}   & \multirow{2}{*}{4,564,519}  & 6 - Malware Detection                      & 2                  \\
                                               &                            &                             & 7 - Application Classification             & 20                 \\
\midrule
\multirow{2}{*}{Cross Platform (Android) 2020~\cite{cp}} & \multirow{2}{*}{66,346}    & \multirow{2}{*}{1,358,292}  & 8 - Application Classification             & 212                \\
                                               &                            &                             & 9 - Country Detection                      & 3                  \\
\midrule
\multirow{2}{*}{Cross Platform (iOS) 2020~\cite{cp}}     & \multirow{2}{*}{34,912}    & \multirow{2}{*}{971,762}    & 10 - Application Classification            & 196                \\
                                               &                            &                             & 11 - Country Detection                     & 3                  \\
\midrule
\multirow{2}{*}{CIRA-CIC-DoHBrw 2020~\cite{doh}}          & \multirow{2}{*}{831,497}   & \multirow{2}{*}{32,962,034} & 12 - DoH Attack Detection                  & 2                  \\
                                               &                            &                             & 13 - DoH Query Method Classification       & 5                  \\
\midrule
\multirow{2}{*}{CIC IoT Dataset 2023~\cite{iot}}          & \multirow{2}{*}{1,163,495} & \multirow{2}{*}{27,738,736}  & 14 - IoT Attack Detection                  & 2                  \\
                                               &                            &                             & 15 - IoT Attack Method Detection           & 7                  \\
\midrule
\multirow{5}{*}{Consecutive Packets from above 7 datasets}                      & \multirow{5}{*}{-} & \multirow{5}{*}{12,786,490} & 16 - Source IP Address Generation            & \multirow{5}{*}{-} \\
 & & & 17 - Destination IP Address Generation & \\
 & & & 18 - Source Port Number Generation & \\
 & & & 19 - Destination Port Number Generation & \\
 & & & 20 - Packet Length Generation & \\
\midrule
Total                      & 2,952,302 & 69,799,192 & -            & - \\
    
\bottomrule
\end{tabular}
\label{tab:stats_of_dataset}
\end{table*}

Over the past decades, machine learning-based methods have been extensively developed for network traffic analysis. Earlier studies~\cite{fs-net,liu2018mampf,bilstm,datanet,deeppacket} primarily assessed the proposed methods on a single dataset encompassing at most 2 classification tasks. Despite achieving notable performance, these studies suffer from disparate data processing pipelines and insufficiently comprehensive evaluations. To address this issue, some recent works~\cite{tscrnn,etbert,yatc} have evaluated performance across multiple datasets and a broader range of classification tasks. For instance, the foundation model ET-BERT~\cite{etbert} was pre-trained on 7 datasets to learn intrinsic representations and then fine-tuned for 5 downstream tasks. However, the data processing techniques in existing works are often customized for themselves.
Moreover, some studies randomly split the extracted packets from the same flow into training and testing dataset for packet-level evaluation, which may cause data leakage since the packets from same flow are strongly correlated. Additionally, certain research practices~\cite{netshare,stan}
only focused on traffic generation for network simulation rather than traffic classification.
In summary, there is a lack of evaluation against a unified and comprehensive network benchmark that encompasses both network traffic classification and generation, making it hard to fairly compare performance.





To address existing limitations, we propose \NetBench, a large-scale
and comprehensive network traffic benchmark covering 7 distinct datasets, featuring 15 classification tasks and 5 generation tasks, as illustrated in Table \ref{tab:stats_of_dataset}. For the classification benchmark, we construct 15 tasks across 7 datasets, including ISCXVPN 2016~\cite{vpn}, ISCXTor 2016~\cite{tor}, USTC-TFC 2016~\cite{ustc},  Cross-Platform~\cite{cp} dataset containing both Android and IOS applications, CIRA-CIC-DoHBrw 2020~\cite{doh}, and CIC IoT Dataset 2023~\cite{iot}. The specific tasks for each dataset are outlined in table \ref{tab:stats_of_dataset}. For the generation benchmark, we devise five tasks tailored to essential header fields~\cite{netshare} at the packet level, including IP addresses (Source/Destination), port number (Source/Destination) and packet length over the same 7 datasets above. This benchmark can help create valid network packets for network simulation.

To prevent data leakage, we first split the original network traffic data into training, validation, and testing sets, from which we extract flows and packets. Then, we anonymize sensitive header fields and employ a unified hexadecimal encoding to standardize different data formats. Moreover, our benchmark provides both flow-level and packet-level evaluations for each dataset, accommodating various input types. To the best of our knowledge, \NetBench~is the first network traffic benchmark that covers a wide range of tasks through a unified data processing method. Furthermore, we evaluate eight SOTA models (including two foundation models, ET-BERT~\cite{etbert} and YaTC~\cite{yatc}) on 15 classification tasks and two generative models on 5 generation tasks using our benchmark. We believe that our benchmark has laid a solid foundation for evaluating network traffic models fairly, which will significantly contribute to the development of foundation models for network traffic.

In summary, the main contributions of this work are three-fold:
\begin{itemize}[noitemsep]
    \item \textbf{Comprehensive Network Traffic Benchmark.} We first create a large-scale and comprehensive network traffic benchmark from 7 distinct datasets with 20 tasks. Our benchmark offers diverse data types for foundation model training and a wide range of downstream tasks for fair evaluation in both network traffic classification and generation.
    \item \textbf{Unified Data Processing.} We introduce a uniform hexadecimal encoding method to process diverse network traffic data with anonymization in sensitive header fields. Our data processing method unifies data pre-processing, data standardization, and data segmentation, which can offer a fair evaluation opportunity and save human labor. 
    \item \textbf{Benchmarking SOTA Methods.} We compare the performance of eight SOTA classifiers (including two foundation models) and two generative models using our benchmark. The evaluation results demonstrate the superiority of foundation models over other approaches in traffic classification. 
\end{itemize}

\section{Related Work}

\textbf{Evaluation on Single Dataset.} Most earlier machine learning-based studies have been evaluated using a single dataset focused on no more than two tasks. For example, FS-Net~\cite{fs-net} assessed its performance on a campus network dataset~\cite{liu2018mampf} with 18 classes by extracting statistical features like packet length as model input. Similarly, BiLSTM\_ATTN~\cite{bilstm} implemented traffic segmentation and unwanted information removal for data processing before evaluation on the ISCXVPN-detection dataset~\cite{vpn}. DataNet~\cite{datanet} employed balanced subsets, byte vectorization, and normalization in its preprocessing steps, but its testing was confined to the ISCX VPN-application task. In addition, DeepPacket~\cite{deeppacket} built upon earlier efforts~\cite{bilstm, datanet} to eliminate irrelevant packets for evaluation on the ISCXVPN dataset. STAN~\cite{stan} applied normalization and one-hot encoding for pre-processing before it was evaluated on a selected subset from UGR'16~\cite{macia2018ugr} dataset across 2 generation tasks. While these approaches have shown promising results, their evaluations were restricted to single dataset and a limited range of tasks. Moreover, the variance in their data processing techniques could result in biased comparisons.

\noindent
\textbf{Evaluation on Multiple datasets.} To achieve a more comprehensive evaluation, recent works have assessed their models on more datasets covering a variety of tasks. For instance, TSCRNN~\cite{tscrnn} was evaluated on 3 public datasets following the pre-processing pipelines established in~\cite{bilstm, datanet}. More recently, Transformer-based foundation models like ET-BERT~\cite{etbert} pre-trained on 6 public and 1 collected dataset to learn network traffic representations and then fine-tuned for 5 classification tasks. However, the construction of their packet-level evaluation datasets through random sampling raises the potential issue of data leakage, as highly correlated packets from the same flow could appear in both the training and testing datasets. Additionally, the foundation model YaTC~\cite{yatc} pre-trained on 4 datasets for better representation and fine-tuned on 5 public datasets to test its generalization ability. Nevertheless, it faces a similar risk of data leakage for packet-level evaluation due to its use of a data processing methodology akin to ET-BERT. Meanwhile, NetShare~\cite{netshare} evaluated its generative performance across six datasets, although it is not designed for traffic classification tasks.


In summary, while current approaches have been assessed across various datasets and tasks, the comprehensiveness and fairness of these evaluations remain questionable due to the variance in data processing approaches. To our best knowledge, there has yet to be an establishment of a large-scale and comprehensive benchmark that enables a fair comparison across different models.


\begin{table}[!t]
    \centering
    \caption{Comparison of network traffic methods in terms of dataset utilization and task diversity. Our benchmark encompasses a total of 20 tasks across 7 datasets.}
    \begin{tabular}{cccc}
    \toprule
    \multirow{2}{*}{\centering Method}     & \multicolumn{1}{c}{\#Employed} & \multicolumn{1}{c}{\#Dataset Size} & \multicolumn{1}{c}{\#Covered} \\
               & \multicolumn{1}{c}{Dataset(s)} & \multicolumn{1}{c}{(flows)} & \multicolumn{1}{c}{Task(s)} \\ \midrule
    FS-Net     & 1        & 956K   & 1               \\
    BiLSTM     & 1        & 47K    & 2               \\
    Datanet    & 1        & 206K   & 1               \\
    DeepPacket & 1        & 206K  & 2               \\
    STAN       & 1        & 348K        & 2               \\
    TSCRNN     & 3        & 420K      & 3               \\
    YaTC       & 5        & 360K        & 4               \\
    ET-Bert    & \textbf{7}        & 376K        & 5               \\
    NetShare   & 6        & 600K        & 13               \\ 
    \midrule
    \NetBench  & \textbf{7}         & \textbf{2952K}       & \textbf{20}              \\ \bottomrule
    \end{tabular}
    \label{tab:diff}
\end{table}
\section{NetBench}
In this section, we create a large-scale and comprehensive benchmark dataset named \NetBench, designed to standardize the evaluation of network traffic analysis models. Our benchmark contains 7 datasets with 20 tasks, including 15 classification tasks and 5 generation tasks. 

\begin{figure*}[t]
    \centering
    \includegraphics[width=0.95\textwidth]{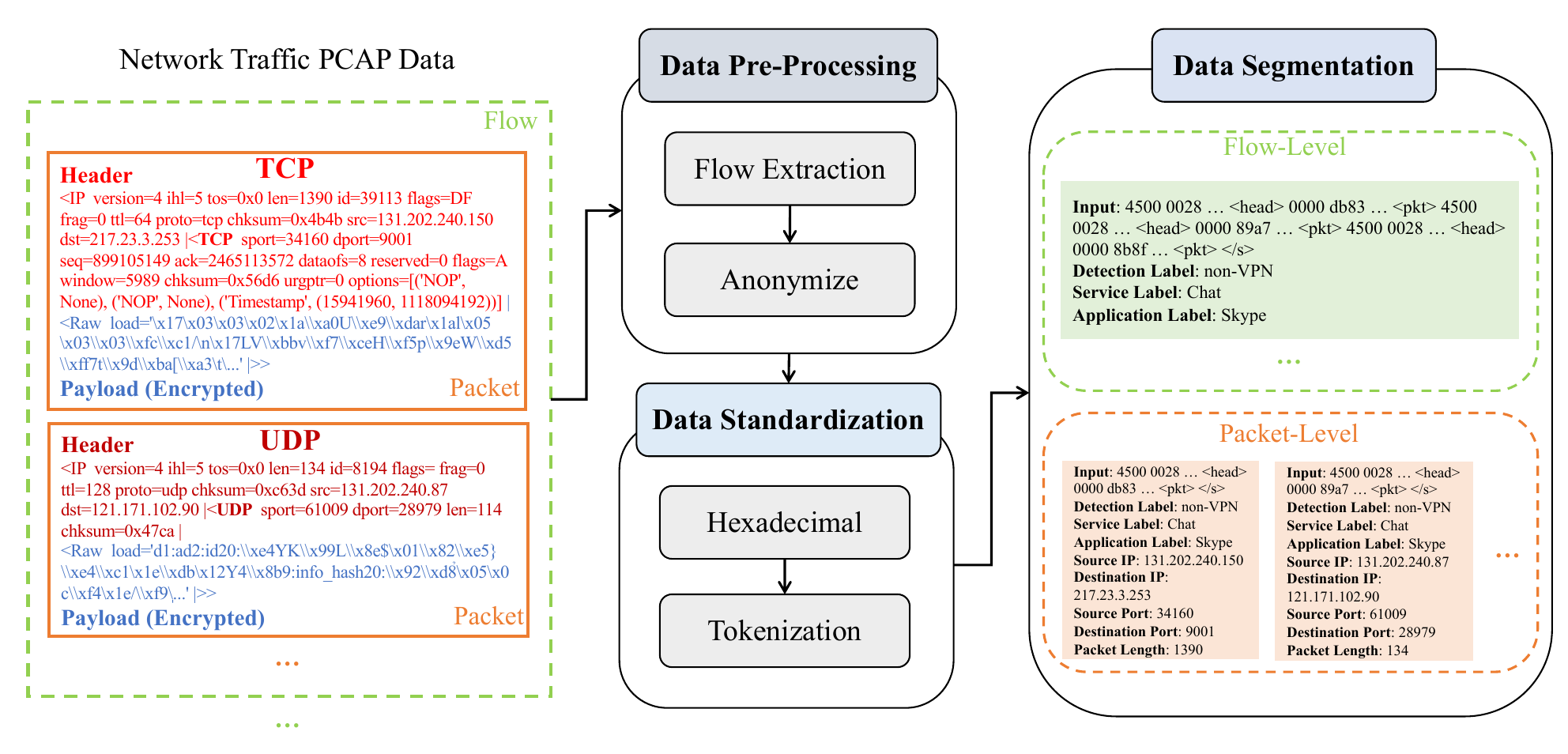}
    \vspace{-0.2in}
    \caption{The overall pipeline of data preparation, consisting of three parts: Data Pre-Processing, Data Standardization, and Data Segmentation. Firstly, we convert flows extracted from network traffic PCAP files into a hexadecimal format. Then, the WordPiece algorithm is employed to segment the hexadecimal data into 4-digit blocks, incorporating specific symbols (\texttt{</s>} for sequence ends, \texttt{<head>} for header separation, and \texttt{<pkt>} for packet demarcation). Lastly, we create two different types of dataset based on flow level and packet level.}
    \label{fig:data_process}
    \vspace{-0.1in}
\end{figure*}

\subsection{Data Preparation}
To create a large-scale and comprehensive benchmark, we first collect raw data samples from 7 publicly available datasets, as detailed in table \ref{tab:stats_of_dataset}. Then, our next step is to convert them into a standardized format as model input. Fig.~\ref{fig:data_process} shows the proposed pipeline of data preparation, which consists of three steps: Data Pre-Processing, Data Standardization, and Data Segmentation.

\begin{itemize}[noitemsep, leftmargin=0.5cm]
    \item \textbf{Data Pre-Processing.} Network traffic data encompasses flows of network packets containing privacy-sensitive information. As network traffic packets are captured and saved in Packet Capture (PCAP) files, we firstly segments PCAP files into distinct set for training, validation, and testing. This strategy ensures that high-correlated packets from the same flow will not be simultaneously present in both training and testing data. This effectively reduces the risk of data leakage. Subsequently, we extract flows containing multiple packets from these PCAP files according to distinct combinations of IP addresses, port numbers and protocols~\cite{splitcap}. To protect data privacy, we further anonymize each packet by masking the source/destination IP addresses and port numbers (replaced with 0). This not only preserves data integrity but also protects data privacy. 
    \item \textbf{Data Standardization.} Since data packets include both plaintext and ciphertext, we need to convert anonymized flows into a hexadecimal format, thereby unifying both header and payloads' formats and reducing the complexity to process network traffic data. This standardization offers a fair and uniform basis for model comparison, benchmarking more accurately among different tasks in network traffic classification and generation. Following the same methodology in~\cite{etbert}, we employ the WordPiece algorithm to segment the hexadecimal data into 4-digit blocks, incorporating specific symbols (\texttt{</s>} for sequence ends, \texttt{<head>} for header separation, and \texttt{<pkt>} for packet demarcation) to standardize the dataset format.
    \item \textbf{Data Segmentation.} To capture the underlying characteristics of network traffic, we offer evaluation on each dataset at both flow level and packet level. Flow-level data is ideal for high-level analysis, allowing for the identification of trends, such as increasing or decreasing volumes of data over time, shifts in the types of traffic, and overall patterns of network use. Conversely, packet-level data enables a detailed examination of packet content and specific network traffic patterns. In addition to classification, we extract IP addresses, port numbers, and packet length for packet-level data as additional labels for generation tasks. By enabling evaluation at both levels, we can comprehensively evaluate the performance of different models in both classification and generation tasks.
\end{itemize}

\subsection{Tasks in \NetBench}
Utilizing the described pipeline on seven collected traffic datasets, we construct a benchmark dataset for network traffic called \NetBench. This dataset encompasses a total of 20 downstream tasks, including 15 traffic classification and 5 traffic generation tasks. Table \ref{tab:stats_of_dataset} provides a detailed summary of \NetBench. To our best knowledge, \NetBench~is the largest and most comprehensive benchmark in network traffic domain. It offers significant benefits for the development of foundational models and enables fair assessment of their performance.

To better understand different tasks, we give two representative examples about traffic classification and generation as follows.

\noindent
\textbf{Classification Tasks:} We utilize the ISCXVPN 2016~\cite{vpn} dataset as an illustrative example. This dataset showcases the emulation of real-world internet behaviors by setting up user accounts to interact with each other on popular applications in VPN-routed or non-VPN scenarios. It involves capturing both VPN-routed and non-VPN internet sessions. There are three tasks: VPN Detection (Task 1) aims to ascertain the use of VPNs; VPN Service Detection (Task 2) seeks to identify 6 specific services, namely P2P, Streaming, Email, Chat, VoIP, and File Transfer; and VPN Application Detection (Task 3) strives to differentiate 17 distinct applications, including Facebook, Skype, YouTube, and more. Similarly, the remaining 12 tasks from other 6 datasets perform different types of classifications.

\noindent
\textbf{Generation Tasks:} The goal of this task is to generate five pivotal header fields: the source IP address (Task 16), destination IP address (Task 17), source port (Task 18), destination port (Task 19), and the packet length (Task 20)~\cite{netshare}. This generation process entails the synthesis of 5 header fields to facilitate the creation of network traffic packets that closely mirrors real-world scenarios. Such generations are crucial to the evaluation of innovative networking hardware and software solutions~\cite{netshare, zhang2022dual}.


\section{Experiments}
In this section, we perform a fair and holistic evaluation of SOTA models on the \NetBench~in both network traffic classification and generation tasks.


\begin{table*}[!thb]
    \centering
    \small
    \caption{Comparison results of traffic classification tasks from tasks 1 to 8 (flow: Flow-level, pkt: Packet-level, AC: Accuracy, F1: F1-score). Foundation models like ET-BERT and YaTC outperform other traditional deep learning methods.}
    \renewcommand\arraystretch{0.25}
    \label{tab:compUnder1}
    \renewcommand\arraystretch{1.2}
    \resizebox{\textwidth}{!}{
    \begin{tabular}{l | c c | c c | c c | c c | c c | c c | c c | c c  }
        \toprule
        \multirow{2}{*}{\centering Method} &
        \multicolumn{2}{c|}{Task 1} &
        \multicolumn{2}{c|}{Task 2} &
        \multicolumn{2}{c|}{Task 3} &
        \multicolumn{2}{c|}{Task 4} &
        \multicolumn{2}{c|}{Task 5} &
        \multicolumn{2}{c|}{Task 6} &
        \multicolumn{2}{c|}{Task 7} &
        \multicolumn{2}{c}{Task 8}\\
        \cline{2-17}
        & AC & F1 & AC & F1 & AC & F1 & AC & F1 & AC & F1 & AC & F1 & AC & F1 & AC & F1\\
        \midrule
        Datanet   (flow)     & 0.9406 & 0.4847 & 0.6918 & 0.1363 & 0.3397 & 0.0298 & 0.9987 & 0.4997 & 0.4981 & 0.0950 & 0.6374 & 0.3893 & 0.2096 & 0.0173 & 0.0945 & 0.0153 \\
Datanet   (pkt)      & 0.8836 & 0.4691 & 0.5993 & 0.1249 & 0.3198 & 0.0285 & 0.9961 & 0.4990 & 0.3196 & 0.0692 & 0.6440 & 0.3917 & 0.0817 & 0.0076 & 0.0207 & 0.0002 \\
Fs-Net (flow)        & 0.9258 & 0.4807 & 0.2930 & 0.3367 & 0.2109 & 0.3088 & 0.9976 & 0.8327 & 0.8203 & 0.6354 & 0.3711 & 0.2801 & 0.8203 & 0.8455 & 0.0708 & 0.0411 \\
BiLSTM\_ATTN   (flow) & 0.9406 & 0.4847 & 0.0057 & 0.0313 & 0.3373 & 0.0484 & 0.9996 & 0.9166 & 0.8833 & 0.5554 & 0.6374 & 0.3893 & 0.0256 & 0.0052 & 0.0045 & 0.0004 \\
BiLSTM\_ATTN   (pkt)  & 0.8836 & 0.4691 & 0.0303 & 0.0098 & 0.3198 & 0.0285 & 0.9961 & 0.4990 & 0.0722 & 0.0192 & 0.6440 & 0.3917 & 0.0329 & 0.0032 & 0.0003 & 0.0000 \\
DeepPacket   (flow)  & 0.9408 & 0.4948 & 0.6918 & 0.1363 & 0.3397 & 0.0298 & 0.9998 & 0.9666 & 0.4981 & 0.0950 & 0.6374 & 0.3893 & 0.3110 & 0.0908 & 0.0484 & 0.0004 \\
DeepPacket   (pkt)   & 0.8836 & 0.4691 & 0.5993 & 0.1249 & 0.3198 & 0.0285 & 0.0039 & 0.0039 & 0.4727 & 0.0917 & 0.6440 & 0.3917 & 0.2638 & 0.0209 & 0.0238 & 0.0002 \\
TSCRNN   (flow)      & 0.9406 & 0.4847 & 0.6918 & 0.1363 & 0.3397 & 0.0298 & 0.9987 & 0.7664 & 0.4470 & 0.1340 & 0.6374 & 0.3893 & 0.0839 & 0.0108 & 0.0243 & 0.0024 \\
TSCRNN   (pkt)       & 0.8836 & 0.4691 & 0.0303 & 0.0098 & 0.3198 & 0.0285 & 0.9961 & 0.4990 & 0.0722 & 0.0192 & 0.6440 & 0.3917 & 0.0329 & 0.0032 & 0.0037 & 0.0000 \\
\midrule
ET-BERT   (flow)     & 0.9964 & 0.9838 & 0.7462 & 0.7110 & 0.5253 & 0.6667 & \textbf{1.0000} & \textbf{1.0000} & 0.9571 & 0.8029 & \textbf{1.0000} & \textbf{1.0000} & 0.9786 & 0.9820 & 0.8463 & 0.6770 \\
ET-BERT   (pkt)      & 0.9902 & 0.9758 & 0.7653 & 0.7631 & 0.5972 & 0.6956 & \textbf{1.0000} & 0.9978 & 0.8469 & 0.5685 & \textbf{1.0000} & \textbf{1.0000} & 0.9539 &      0.9530 & \textbf{0.9687} & \textbf{0.8753} \\
YaTC   (flow)        & 0.9974 & 0.9880 & 0.7805 & 0.7083 & 0.5991 & 0.7090 & \textbf{1.0000} & \textbf{1.0000} & \textbf{0.9739} & \textbf{0.8512} & \textbf{1.0000} & \textbf{1.0000} & \textbf{0.9936} & \textbf{0.9949} & 0.9161 & 0.8228 \\
YaTC   (pkt)         & \textbf{0.9984} & \textbf{0.9961} & \textbf{0.8073} & \textbf{0.8260} & \textbf{0.6458} & \textbf{0.7837} & 0.9998 & 0.9896 & 0.9601 & 0.8297 & \textbf{1.0000} & \textbf{1.0000} & 0.9850 & 0.9874 & 0.9519 & 0.8462 \\
        \bottomrule
    \end{tabular}
    }
\end{table*}

\begin{table*}[!thb]
    \centering
    \small
    \caption{Comparison results of traffic classification tasks from tasks 9 to 15 (flow: Flow-level, pkt: Packet-level, AC: Accuracy, F1: F1-score). Foundation models like ET-BERT and YaTC outperform other traditional deep learning methods.}
    \renewcommand\arraystretch{0.25}
    \label{tab:compUnder2}
    \renewcommand\arraystretch{1.2}
    \resizebox{\textwidth}{!}{
    \begin{tabular}{l | c c | c c | c c | c c | c c | c c | c c}
        \toprule
        \multirow{2}{*}{\centering Method} &
        \multicolumn{2}{c|}{Task 9} &
        \multicolumn{2}{c|}{Task 10} &
        \multicolumn{2}{c|}{Task 11} &
        \multicolumn{2}{c|}{Task 12} &
        \multicolumn{2}{c|}{Task 13} &
        \multicolumn{2}{c|}{Task 14} &
        \multicolumn{2}{c}{Task 15}\\
        \cline{2-15}
        & AC & F1 & AC & F1 & AC & F1 & AC & F1 & AC & F1 & AC & F1 & AC & F1 \\
        \midrule
        Datanet   (flow)     & 0.7795 & 0.2920 & 0.0481 & 0.0005 & 0.7717 & 0.7781 & 0.8261 & 0.4524 & 0.2928 & 0.0906 & 0.9768 & 0.4941 & 0.0250 & 0.0081 \\
Datanet   (pkt)      & 0.8112 & 0.2986 & 0.0429 & 0.0004 & 0.3788 & 0.1832 & 0.8187 & 0.4502 & 0.2907 & 0.0901 & 0.0078 & 0.0078 & 0.0044 & 0.0015 \\
Fs-Net (flow)        & 0.5273 & 0.2314 & 0.1094 & 0.0638 & 0.1758 & 0.0997 & 0.4023 & 0.2869 & 0.1855 & 0.0626 & 0.9023 & 0.7369 & 0.6680 & 0.5481 \\
BiLSTM\_ATTN   (flow) & 0.1152 & 0.0689 & 0.0029 & 0.0001 & 0.3860 & 0.1857 & 0.8261 & 0.4524 & 0.2609 & 0.0952 & 0.9768 & 0.4941 & 0.0579 & 0.0466 \\
BiLSTM\_ATTN  (pkt)  & 0.0988 & 0.0600 & 0.0005 & 0.0000 & 0.3788 & 0.1832 & 0.8187 & 0.4502 & 0.3851 & 0.1658 & 0.9922 & 0.4980 & 0.0044 & 0.0015 \\
DeepPacket   (flow)  & 0.7795 & 0.2920 & 0.0481 & 0.0005 & 0.3660 & 0.1786 & 0.8261 & 0.4524 & 0.5332 & 0.1391 & 0.0232 & 0.0227 & 0.3435 & 0.0852 \\
DeepPacket   (pkt)   & 0.8112 & 0.2986 & 0.0429 & 0.0004 & 0.3788 & 0.1832 & 0.8187 & 0.4502 & 0.2907 & 0.0901 & 0.0078 & 0.0078 & 0.0044 & 0.0015 \\
TSCRNN   (flow)      & 0.7795 & 0.2920 & 0.0284 & 0.0051 & 0.3860 & 0.1857 & 0.8261 & 0.4524 & 0.2928 & 0.0906 & 0.9768 & 0.4941 & 0.0250 & 0.0081 \\
TSCRNN   (pkt)       & 0.0988 & 0.0600 & 0.0005 & 0.0000 & 0.3788 & 0.1832 & 0.8187 & 0.4502 & 0.2907 & 0.0901 & 0.9922 & 0.4980 & 0.0044 & 0.0015 \\
\midrule
ET-BERT   (flow)     & 0.9762 & 0.9418 & 0.7705 & 0.7426 & \textbf{0.9881} & \textbf{0.9866} & \textbf{1.0000} & \textbf{1.0000} & \textbf{0.9998} & 0.9980 & 0.9881 & \textbf{0.8535} & \textbf{0.8809} & \textbf{0.8329} \\
ET-BERT   (pkt)      & 0.9911 & 0.9785 & 0.9435 & 0.9313 & 0.9736 & 0.9740 & 0.9970 & 0.9948 & 0.9701 & 0.9316 & \textbf{0.9957} &  0.8526  & 0.8222 &    0.7804 \\
YaTC   (flow)        & 0.9927 & 0.9782 & 0.7531 & 0.6957 & 0.9736 & 0.9705 & \textbf{1.0000} & \textbf{1.0000} & 0.9990 & \textbf{0.9989} & 0.9825 & 0.8356 & 0.8618 & 0.7316 \\
YaTC   (pkt)         & \textbf{0.9981} & \textbf{0.9933} & \textbf{0.9576} & \textbf{0.9407} & 0.9868 & \textbf{0.9866} & \textbf{1.0000} & \textbf{1.0000} &  0.9682	& 0.7499 & 0.9936 & 0.8440 &  0.8639	& 0.6254 \\
        \bottomrule
    \end{tabular}
    }
\end{table*}
\subsection{Experimental Settings}
\textbf{SOTA Models}.
We assess the performance of 8 major SOTA open-source models on~\NetBench~to ensure a comprehensive and unbiased comparison. Specifically, they include 7 classification models, DataNet~\cite{datanet}, FS-Net~\cite{fs-net}, BiLSTM\_ATTN~\cite{bilstm}, DeepPacket~\cite{deeppacket}, TSCRNN~\cite{tscrnn}, ET-BERT~\cite{etbert}, YaTC~\cite{yatc}, as well as two competitive generative models, STAN~\cite{stan} and Netshare~\cite{netshare}. Note that, ET-BERT and YaTC are foundation models which could be pre-trained and then fine-tuned on different classification tasks. The detailed experimental settings of each model are described below. For FS-Net, we follow its setting that only evaluate its performance in classification tasks at flow-level. For DataNet, DeepPacket, TSCRNN, and BiLSTM\_ATTN, we assess their performance on all classification tasks at both flow level and packet level. Regarding ET-Bert and YaTC, we employ the released pre-trained weights and fine-tune them with the entirety of prepared training dataset. For generative models, we following the same setting in prior work~\cite{netshare} to randomly sample consecutive packets from each dataset, ensuring a consistent evaluation for future studies.


\noindent
\textbf{Dataset.} We split our benchmark into training, validation, and testing data with a ratio of 8:1:1 respectively.

\noindent
\textbf{Evaluation metrics.} For assessing SOTA models in traffic understanding tasks, we employ two principal metrics: Accuracy (AC) and Macro F1 Score (F1). To evaluate the traffic generation performance, we utilize the Jensen-Shannon Divergence (JSD) and Total Variation Distance (TVD). JSD quantifies the similarity of two probability distributions, indicating their shared information, while TVD measures the largest difference between two probability distributions, capturing the maximum discrepancy.

\subsection{Evaluation Results}
\textbf{Classification Tasks.} Table~\ref{tab:compUnder1} and~\ref{tab:compUnder2} illustrate the comparison results of the 8 SOTA models for traffic classification using \NetBench. From the evaluation results, we can conclude with three main insights: (1) Traditional deep learning methods, such as DataNet, DeepPacket, FS-Net, TSCRNN and BiLSTM\_ATTN, exhibit limitations in generalizing to new tasks, with a noticeable tendency to bias classifications towards dominant classes. This is evidenced by the f1-score that is consistently below 0.5 with low recall score due to data imbalance. (2) Foundation models like ET-BERT and YaTC significantly outperform these traditional approaches, showcasing their superior prediction accuracy and generalization ability. (3) Although a flow with multiple packets inherently contains richer information compared to a single packet, foundation models trained at flow level do not surpass those trained on packet-level in tasks 3, 8, and 10. This is attributed to the constraints on the input length for foundation models as the input flows exceeding a specified length will be truncated. 


\begin{table*}[thb]
    \centering
    \small
    \caption{Evaluation results of generation tasks (Task 16 - 20) using STAN and NetShare. Netshare exhibits good performance in generating IP addresses and port numbers on most datasets while STAN performs better in generating Packet Length.}
    \renewcommand\arraystretch{0.36}
    \label{tab:compGen}
    \renewcommand\arraystretch{1.1}
    \resizebox{\textwidth}{!}{
    \begin{tabular}{l | l | cc | cc | cc | cc | cc }
    \toprule
\multirow{2}{*}{\centering Dataset}        & \multirow{2}{*}{\centering Method} & \multicolumn{2}{c|}{16 - Source IP}              & \multicolumn{2}{c|}{17 - Destination IP}         & \multicolumn{2}{c|}{18 - Source Port}            & \multicolumn{2}{c|}{19 - Destination Port}       & \multicolumn{2}{c}{20 - Packet Length}          \\
                                \cline{3-12} &  &  \multicolumn{1}{c}{JSD} & \multicolumn{1}{c|}{TVD} & \multicolumn{1}{c}{JSD} & \multicolumn{1}{c|}{TVD} & \multicolumn{1}{c}{JSD} & \multicolumn{1}{c|}{TVD} & \multicolumn{1}{c}{JSD} & \multicolumn{1}{c|}{TVD} & \multicolumn{1}{c}{JSD} & \multicolumn{1}{c}{TVD} \\
\midrule
\multirow{2}{*}{ISCXVPN 2016}                    & STAN                                        & \textbf{0.1130}            & 0.4107           & \textbf{0.0850}              & \textbf{0.3407}              & \textbf{0.0186}            & \textbf{0.1220}           & \textbf{0.0247}               & \textbf{0.1791}               & \textbf{0.0767}              & \textbf{0.3405 }            \\
                                                 & NetShare                                    & 0.1218            & \textbf{0.3513}           & 0.1269              & 0.3603              & 0.1622            & 0.4966           & 0.1494               & 0.4755               & 0.5451              & 0.9011             \\
\midrule
\multirow{2}{*}{ISCXTor   2016}                  & STAN                                        & 0.2089            & 0.5505           & 0.2132              & 0.5615              & 0.2163            & 0.5658           & 0.1998               & 0.5225               & \textbf{0.2173}              & \textbf{0.5665}             \\
                                                 & NetShare                                    & \textbf{0.0168}            & \textbf{0.0880}           & \textbf{0.0338}              & \textbf{0.1263}              & \textbf{0.0736}            & \textbf{0.2190}           & \textbf{0.0711}               & \textbf{0.2269}               & 0.3091              & 0.7004             \\
\midrule
\multirow{2}{*}{USTC-TFC   2016}                 & STAN                                        & 0.4421            & 0.8134           & 0.4226              & 0.7855              & 0.4070            & 0.7829           & 0.4767               & 0.8515               & \textbf{0.3235}              & \textbf{0.6952}             \\
                                                 & NetShare                                    & \textbf{0.0363}            & \textbf{0.2148}           & \textbf{0.0445}              & \textbf{0.2205}              & \textbf{0.2268}            & \textbf{0.5899}           & \textbf{0.1144}               & \textbf{0.4091}               & 0.4908              & 0.8598             \\
\midrule
\multirow{2}{*}{Cross   Platform (Android) 2020} & STAN                                        & 0.2446            & 0.5834           & 0.4743              & 0.8498              & 0.4089            & 0.7850           & 0.3727               & 0.7479               & \textbf{0.2196}              & \textbf{0.5845}             \\
                                                 & NetShare                                    & \textbf{0.0572}            & \textbf{0.2909}           & \textbf{0.0155}              & \textbf{0.1177}              & \textbf{0.0520}            & \textbf{0.2177}           & \textbf{0.0323}               & \textbf{0.2216}               & 0.4266              & 0.8008             \\
\midrule
\multirow{2}{*}{Cross   Platform (iOS) 2020}     & STAN                                        & 0.3092            & 0.6537           & 0.5368              & 0.9093              & 0.4594            & 0.8391           & 0.4393               & 0.8181               & \textbf{0.2538}              & \textbf{0.6002}             \\
                                                 & NetShare                                    & \textbf{0.0588}            & \textbf{0.2665}           & \textbf{0.0261}              & \textbf{0.1717}              & \textbf{0.0499}            & \textbf{0.2176}           & \textbf{0.0217}               & \textbf{0.1592}               & 0.3981              & 0.7748             \\
\midrule
\multirow{2}{*}{CIRA-CIC-DoHBrw   2020}          & STAN                                        & 0.3710            & 0.7382           & 0.2823              & 0.5867              & 0.2351            & 0.5336           & \textbf{0.0960}               & \textbf{0.2511}               & \textbf{0.0313}              & \textbf{0.0875}             \\
                                                 & NetShare                                    & \textbf{0.3548}            & \textbf{0.7113}           & \textbf{0.2481}              & \textbf{0.5828}              & \textbf{0.0359}            & \textbf{0.1749}           & 0.0963               & 0.3605               & 0.4806              & 0.8576             \\
\midrule
\multirow{2}{*}{CIC   IoT Dataset 2023}          & STAN                                        & 0.3094            & 0.6537           & 0.5370              & 0.9093              & 0.4596            & 0.8391           & 0.4394               & 0.8182               & \textbf{0.3101}              & \textbf{0.6548}             \\
                                                 & NetShare                                    & \textbf{0.0452}            & \textbf{0.2598}           & \textbf{0.0111}              & \textbf{0.0817}              & \textbf{0.0666}            & \textbf{0.2636}           & \textbf{0.0370}               & \textbf{0.2126}               & 0.4199              & 0.7957                               \\
\bottomrule
\end{tabular}
}
\end{table*}

\noindent
\textbf{Generation Tasks.} We also compare the generation performance of two open-sourced generative models, STAN and NetShare, as shown in Table \ref{tab:compGen}. It can be observed that NetShare performs well in generating IP addresses and port numbers, while STAN achieves better performance on packet length generation. The main reason is that NetShare designs bitwise encoding~\cite{netshare} for IP addresses and IP2Vec encoding~\cite{ip2vec} for port numbers, which helps generating these fields accurately. Conversely, STAN integrates both CNN and mixture density neural layers to capture joint distribution of packet length effectively, explaining its superior performance in generating packet lengths.

\section{Conclusion}
We introduced \NetBench, a large-scale and comprehensive benchmark for network traffic analysis, which addressed the critical issue of fair evaluation and comparison among different methods. Characterized by its comprehensive design, \NetBench~included a total of 20 evaluation tasks across 7 datasets through a unified data processing approach. Furthermore, we evaluated some SOTA models on our benchmark. The experimental results underscored the benefits of foundation models, which demonstrated superior performance across a wide range of classification tasks. These observations highlighted the significant potential of foundation models to revolutionize network traffic analysis. For traffic generation, very few foundation models have investigated in this field. This remains to explore in the future as foundation models have the potential to excel in generation tasks as they do in classification~\cite{lens}. Besides, longer input length is needed for foundation models to better understand the increasingly complex network traffic. In short, future research could leverage \NetBench~to train foundation models in a convenient and straightforward manner for a variety of downstream network traffic classification and generation tasks. We believe that our benchmark will promote advancements in foundation models for network traffic through fair and comprehensive comparisons.



\bibliographystyle{abbrv}
\bibliography{reference.bib}


\end{document}